\documentclass[twocolumn,preprint]{aastex62}
\usepackage[utf8x]{inputenc}
\usepackage{amsmath,amsfonts,amssymb,mathrsfs,bm,enumerate,graphicx}
\usepackage[T1]{fontenc}
\usepackage[all]{hypcap}
\usepackage{ulem}
\usepackage{xcolor}

\usepackage{ae,aecompl}
\hypersetup{breaklinks,colorlinks,citecolor=blue,linkcolor=magenta}

\newcommand{\fargo}{{\footnotesize FARGO3D}}

\newcommand{\ppderiv}[2]{\frac{\partial #1}{\partial #2}}

\newcommand{\avg}[1]{\left\langle #1 \right\rangle}

\newcommand{\be}{\begin{eqnarray}}
\newcommand{\ee}{\end{eqnarray}}

\newcommand\as{\bgroup\markoverwith{\textcolor{red}{\rule[0.5ex]{8pt}{1.5pt}}}\ULon}
\newcommand\ds{\bgroup\markoverwith{\textcolor{teal}{\rule[0.5ex]{8pt}{1.5pt}}}\ULon}
\newcommand{\bmat}{\begin{pmatrix}}
\newcommand{\emat}{\end{pmatrix}}

\newcommand{\rev}[1]{#1}

\renewcommand{\vec}[1]{{\bf #1}}

\defcitealias{DLL}{DLL}

\shorttitle{Outward Migration of Super-Jupiters}
\shortauthors{Dempsey, Mu\~{n}oz, \& Lithwick}

\begin{document}

\title{
Outward Migration of Super-Jupiters
}
\author[0000-0001-8291-2625]{Adam M. Dempsey}
\affiliation{Theoretical Division, Los Alamos National Laboratory, Los Alamos, NM 87545, USA}
\author[0000-0003-2186-234X]{Diego J. Mu\~{n}oz}
\affiliation{Center for Interdisciplinary Exploration and Research in Astrophysics (CIERA)
and
Department of Physics and Astronomy
Northwestern University \\
2145 Sheridan Road
Evanston, IL 60208
USA}
\affiliation{Facultad de Ingenier\'ia y Ciencias, Universidad Adolfo Ib\'a\~nez, Av.\ Diagonal las Torres 2640, Pe\~nalol\'en, Santiago, Chile}
\author[0000-0003-4450-0528]{{Yoram Lithwick}}
\affiliation{Center for Interdisciplinary Exploration and Research in Astrophysics (CIERA)
and
Department of Physics and Astronomy
Northwestern University \\
2145 Sheridan Road
Evanston, IL 60208
USA}

\correspondingauthor{Adam M. Dempsey}
\email{adempsey@lanl.gov}

\begin{abstract}

Recent simulations show that  giant planets of about one Jupiter mass migrate inward at a rate that differs from the Type II prediction. 
 Here we show that at higher masses, planets migrate outward.
Our result differs from previous ones
because
of our  longer simulation times, lower viscosity, and our boundary conditions
 that allow the disk to reach viscous steady state.
  We show that, for planets on circular orbits, the transition 
  from inward to outward migration coincides with the known transition
  from circular to eccentric disks that occurs
  for planets more massive than a few Jupiters.  
  In an eccentric disk, the torque on the outer disk weakens
  due to two effects: the planet launches  weaker waves, and 
  those waves travel further before damping. As a result, the torque
  on the inner disk dominates, and the planet pushes itself outward. 
    Our results suggest that the many super-Jupiters observed by direct-imaging
   at large distances from the star may have gotten there by outward migration. 

\end{abstract}

\keywords{planet-disk interactions, protoplanetary disks, circumstellar disks, planet formation}

\section{Introduction}

It has long been thought that gap-opening planets migrate by 
Type II migration \citep{Ward.1997,Armitage.2010,Kley.2012b}. In  Type II, it is hypothesized  that
planets open gaps that are empty of gas. As a result, 
gas does not cross a planet's orbit, and the planet 
is forced to migrate in lockstep with 
the disk's inward accretion flow. 
But  this hypothesis is at odds  with  hydrodynamical simulations, 
which find that inward gas flow is largely unimpeded by a deep  gap \citep{Lubow.1999,Crida.2006,Fung.2014,Duffell.2014,Durmann.2015,Kanagawa.2017,Robert.2018}.
Furthermore, recent simulations  show that  the migration rate of
Jupiter-mass planets differs from the Type II prediction
\citep[][hereafter \citetalias{DLL}]{Durmann.2015,Kanagawa.2018,DLL}.
And as shown in Figure 12 of \citetalias{DLL}, the rates
found by different groups agree with each other.
 Theoretically, the reason for the failure of Type II 
is   understood (see Section \ref{sec:pd}).

But the question of what happens for planets more massive than Jupiter has not  been adequately 
addressed.
One might expect that sufficiently massive planets clear deep enough gaps to satisfy the criterion of Type II migration.
However, once a planet's mass exceeds around twice that of Jupiter, it excites the disk's eccentricity \citep{Kley.2006,Regaly.2010,Teyssandier.2016,Teyssandier.2017}, 
and the eccentricity of the gas will affect how the planet is torqued.
This effect has not been explored in viscous steady state, and the resulting long-term migration rate is thus poorly constrained.

A change in migration direction 
should occur for very massive secondaries, since it is now known
 that binary stars of near-equal mass ratio expand while embedded in a disk \citep{Munoz.2019,Munoz.2020,Moody.2019,Duffell.2020}.
Indeed, such a transition was found by \citet{Duffell.2020} to occur in the brown dwarf regime, albeit using viscosities and scale heights larger than are typically expected in protoplanetary disks. 
In this paper,  we show that 
 the migration transition depends on viscosity and scale height, and that
 in more realistic disks, the transitional companion
mass is around twice the mass of Jupiter. This transition coincides with the development of substantial eccentricity in the disk.

\subsection{Theory: Beyond Type II}
\label{sec:pd}

A planet embedded in a disk excites
waves \citep{Goldreich.1979}. 
These waves carry angular momentum, and therefore torque the disk where they damp,
and thereby alter the disk's surface density profile. 
Furthermore, the waves' angular momentum comes at the expense of the planet's, and
so the planet must migrate.
\cite{Goldreich.1980}, \cite{Artymowicz.1993}, and \cite{Ward.1997}
 worked
out how much angular momentum is excited in  waves, and their theory adequately predicts
what is found in simulations, even for very deep gaps and Jupiter-mass planets---provided
the surface density profile is known \citepalias[see e.g., the comparison in Figure 14 of][]{DLL}.

There is, however, an important missing component:  {\it how far do waves travel 
before damping?}  The answer  is crucial, because even if the waves' angular momentum
 is known, one can only determine how the waves affect the disk's (azimuthally-averaged)
surface density profile if one knows {\it where} that angular momentum is deposited.  
The surface density profile, in turn, determines the amplitude of the excited waves---and
hence the angular momentum carried by them. 
\cite{Ward.1997} 
answered this question
by assuming that waves damp immediately upon being excited.
With this assumption, gaps are exponentially deep  for planet masses slightly greater than the gap-opening threshold \citep{Syer.1995,Ward.1997}.
Therefore, disk material is strongly prevented from crossing the planet's orbit, and the
planet is  locked in the disk---a.k.a. the Type II paradigm.  

The assumption of zero damping length is unfounded, as recognized
 by \citet[][Sec. IV-b]{Ward.1997}, and  many others
\citep[e.g.,][]{Lunine.1982,Goodman.2001,Rafikov.2002b,Duffell.2015,Kanagawa.2015}.
If the waves travel before damping, the predicted gap becomes {\it much} shallower
than the exponential depth at zero damping length
(\citealt{Crida.2006}; \citealt{Duffell.2015}; \citealt{Kanagawa.2015}; \citealt{Ginzburg.2018}; \citetalias{DLL}).
This is true even if the damping length is modest---of order a scale height. 
The fact that simulated gaps are not exponentially deep, and the reason
for the failure of Type II, is therefore simply that waves travel before damping.\footnote{\citet{Scardoni.2020} present simulation results in which planets migrate close to 
 the predicted Type II rate,
 after transients have died away.   They therefore claim 
 that Type II is correct.  
 But their final migration rates
 are only close to the predicted value\rev{---and, in fact,  are in agreement with the prior  simulation results that show a clear difference from Type II (\citealt{Durmann.2015};\citealt{Kanagawa.2018};\citetalias{DLL};\citealt{Kanagawa.2020}}.
 More tellingly, they plot the  mass flow through the gap, and find  it does not vanish, in 
 contradiction with the basic postulate of Type II. 
 }
In order to accurately determine where 
in the disk the waves damp, and thereby 
 obtain the gap depth and the torques on the planets, one may build a theory, 
 as has been done for low-mass planets \citep[e.g.,][]{Goodman.2001}.
 But for near-Jupiter-mass planets, which excite strongly nonlinear waves, the
 theory has thus far proven intractable.
 Therefore, we turn to  hydrodynamical simulations.

\section{Simulations}

We run  2D  hydrodynamical simulations
that are very similar to those in \citetalias{DLL}. 
But we focus now on super-Jupiters.
 The disk is fed
at large radii with constant mass flux ($\dot{M}$), and 
 is evolved long enough that it reaches viscous steady state.
The planet is fixed on a circular orbit with radius $r_p$ and orbital frequency $\Omega_p$.
We assume that the disk has a sufficiently low mass 
  that it is appropriate to treat the planet's orbit as fixed when finding
  the disk's steady-state structure. 
 As shown in Section \ref{sec:sum}
that assumption restricts the disk's mass to being less than the planet's.

We use the staggered-mesh GPU accelerated code {\tt FARGO3D} \citep{Benitez-Llambay.2016}.
The disk  feels the gravity of both planet and star, as well as
  pressure and viscosity. 
The disk is locally isothermal with  sound speed  
$c_s=h\Omega r$, 
 where  $h$ is the spatially-constant disk aspect ratio, $r$ is distance to the star, and $\Omega$ is the Keplerian orbital frequency. 
 We adopt the \citet{Shakura.1973} $\alpha$-viscosity model, 
 setting the kinematic shear viscosity to $\nu=\alpha c_s^2/\Omega$
 with spatially constant $\alpha$.

The code is run in the frame centered on the star and co-rotating with the planet.
 We therefore include 
 the indirect potential due to the acceleration of the star by the planet.
Further details regarding  the indirect potential  are in Appendix \ref{sec:app_ind}.

\rev{We divide the mesh into nearly square cells, with  dimensions $\Delta \ln r = \Delta\phi=$ constant throughout the domain.}
For the bulk of our simulations, we 
use a resolution of 8 cells per scale height.
 In Appendix \ref{sec:app_conv} (Figure \ref{fig:conv}), 
we show with an example simulation that this grid-size is adequate for determining the total
torque injected by the planet.

The principal result of each simulation is the  torque on the disk in steady-state, $\Delta T$, 
which is the sum of the (positive) torque on the outer disk and the (negative) torque on the inner disk:
\be 
\Delta T = -\int_{{\rm disk}} \Sigma \ppderiv{\Phi_p}{\phi} r dr d\phi , 
\ee
where $\Phi_p$ is the planet's gravitational potential, and $\Sigma$ is the gas surface density.
The {\it normalized torque} is $\Delta T/ ({\dot{M}\ell_p})$, where $\ell_p = r_p^2 \Omega_p$.
It is a function only of
our
 three basic parameters---$\alpha$, $h$, and the planet-star mass ratio ($q$); in particular, it is independent of
$\dot{M}$ and hence of disk mass.

\subsection{Boundary Conditions} \label{sec:bc}

The main difference between the simulations in \citetalias{DLL} and those done by other
groups is the boundary conditions.
Conventionally, the disk is fixed to its initial profile
at the boundaries. 
But that does not allow the disk to relax to a steady-state profile that is independent of the boundary locations.
Instead, at our outer boundary  the disk is fed with a constant $\dot{M}$ by using
the analytic solution for a steady-state disk 
beyond the zone where
the waves have damped.
That solution permits a  pile-up or deficit of gas beyond the planet's orbit.
And at the inner boundary, 
 material is drained without injecting angular momentum, which allows the  disk there
 to reach the  surface density profile of a free steady disk, which has $\dot{M} =3\pi\nu\Sigma$.
In practice, this  is done by setting $\nu\Sigma$ to be constant across the inner boundary 
\citep[see also the discussion of the inner boundary condition in][]{Dempsey.2020}.
Our boundary conditions are valid whether or not the Type II hypothesis is correct. If it is correct, i.e., if no material
crosses the planet's orbit,
then the inner disk in our simulations
would drain away, and the outer pile-up would continually grow.
Instead, we find the disk reaches viscous steady state, with
a constant flow past the planet.

Our boundary conditions here differ from  \citetalias{DLL} in one significant way:
we 
enforce the outer boundary condition in the barycentric frame (see Appendix \ref{sec:app_bc} for details).
If we instead use the stellocentric frame, we find that the indirect terms
generate an artificial eccentricity near the outer boundary---an effect that was less
pronounced in \citetalias{DLL} because of the smaller mass planets considered there. 
In many of these new simulations, the disks are found to be eccentric.  But in all cases, the eccentricity
near the outer boundary is sufficiently small that it is appropriate to treat orbits there as circular around the barycenter
(e.g., Figure \ref{fig:profs} below).

\rev{For most of our simulations, we place the} computational boundaries at $r_{\rm in} = 0.3 r_p$ and $r_{\rm out} = 12 r_p$, where $r_p$ is the planet's orbital radius.
To prevent wave reflections near the inner boundary, we place  wave-killing region between $r=[0.3r_p, 0.4r_p]$\footnote{\rev{Our simulations with highest $h$ have a smaller inner boundary: $r_{\rm in}=0.1r_p$, and wave-killing up to 
$r=0.2r_p$, because wave deposition extends further at high $h$.} \label{foot:2}}.
Our method of wave-killing preserves angular momentum \citepalias{DLL}, and so  captures all of the torque
injected by the planet in the computational domain.
Wave-killing is unnecessary at the distant outer boundary, because waves dissipate before reaching it.

\section{Results}
\subsection{Planet Migration and Disk Eccentricity} \label{sec:res}
\begin{figure}
    \centering
    \includegraphics[trim={0.3cm 0.3cm 0.22cm 0},clip,width=.44\textwidth]{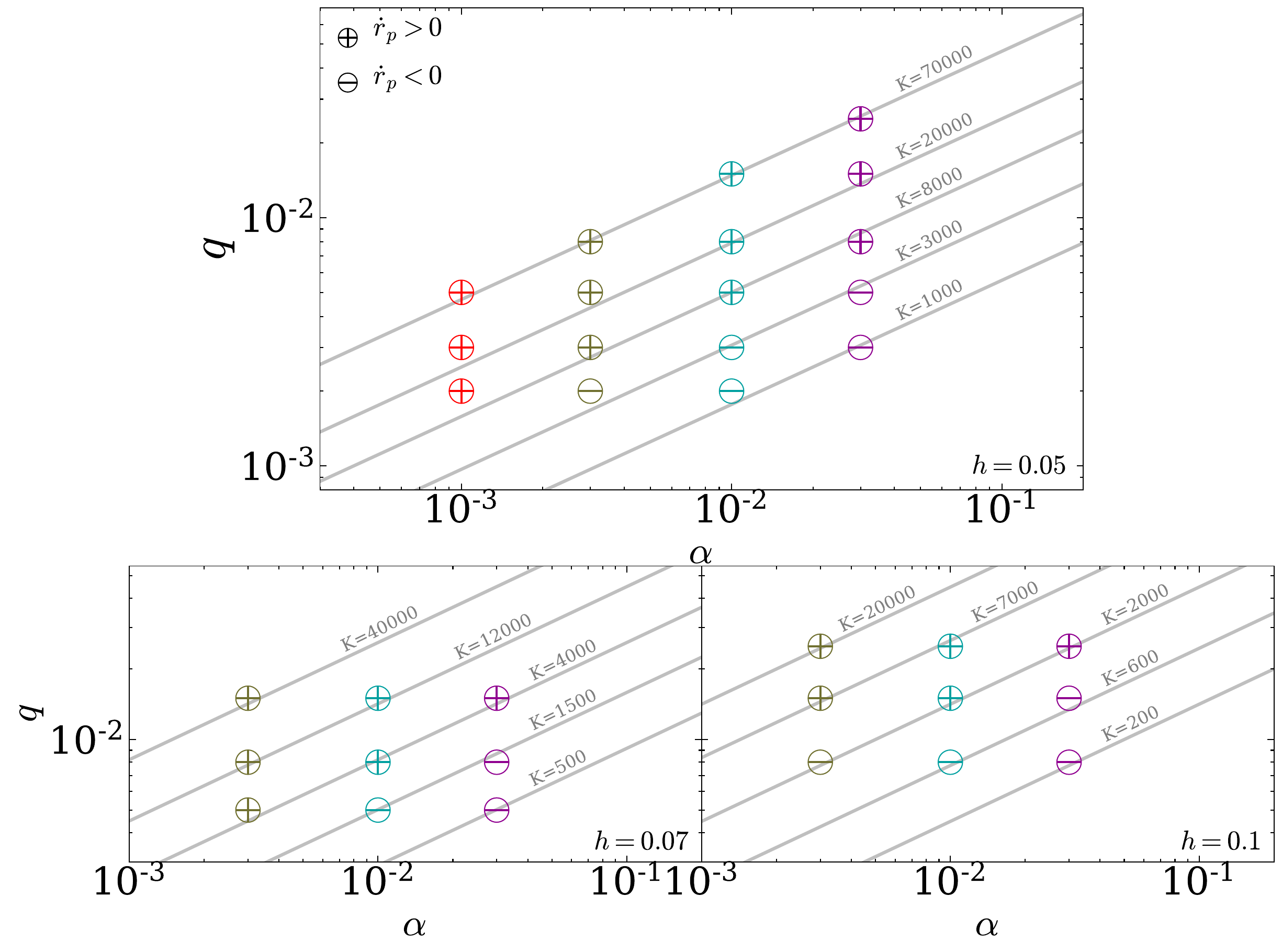}
    \caption{
    Parameters covered by our simulations. 
   We denote the resulting direction of planet migration with a plus sign for outward migration, and 
   a minus sign for inward. 
   Diagonal lines have constant 
     $K = q^2/(\alpha h^5)$.
}
    \label{fig:param}
\end{figure}

We run simulations with the $q$, $\alpha$, and $h$  values shown  in
Figure \ref{fig:param}.
Each simulation is run until the time- and azimuthally-averaged 
$\dot{M}(r)$ profile is spatially constant to within $<10\%$
 (see Appendix \ref{sec:app_conv}). 
In many of the simulations, $\Delta T$ does not reach a constant, but instead 
 varies quasi-periodically in steady state (see Figure \ref{fig:conv}).
As discussed below, the quasi-periodicity is associated with 
the disk being eccentric and apsidally precessing.

Convergence typically takes up to seven
viscous times, evaluated at $r_p$.
The simulations are computationally expensive both because we run for a number of 
viscous times, and because some disks become eccentric, which forces 
the time-step to be smaller.
For example, on a single P100 GPU, one viscous time at $r_p$ 
takes $\sim 20$
 days of wall-time for our lowest $\alpha$ simulations.

In Figure \ref{fig:param}, we plot a plus or minus sign to indicate the direction of the planet's migration.
At each $h$, the transition between inward and outward migration is roughly determined
by the value of   $K \equiv q^2/(\alpha h^5)$, with the critical value of $K$ depending on $h$.
We note that $K$ is equal to the ratio of the one-sided  torque in the absence of a gap ($\propto q^2/h^3$) to the viscous torque ($\propto \alpha h^2$) 
\citep{Lin.1986,Duffell.2013,Kanagawa.2015}, and is found in simulations to set the gap depth 
when $q\lesssim 10^{-3}$ \citep[][]{Duffell.2013,Fung.2014,Kanagawa.2017}.

\begin{figure*}
    \centering
    \includegraphics[trim={0.3cm 0.3cm 0.22cm 0},clip,width=.68\textwidth]{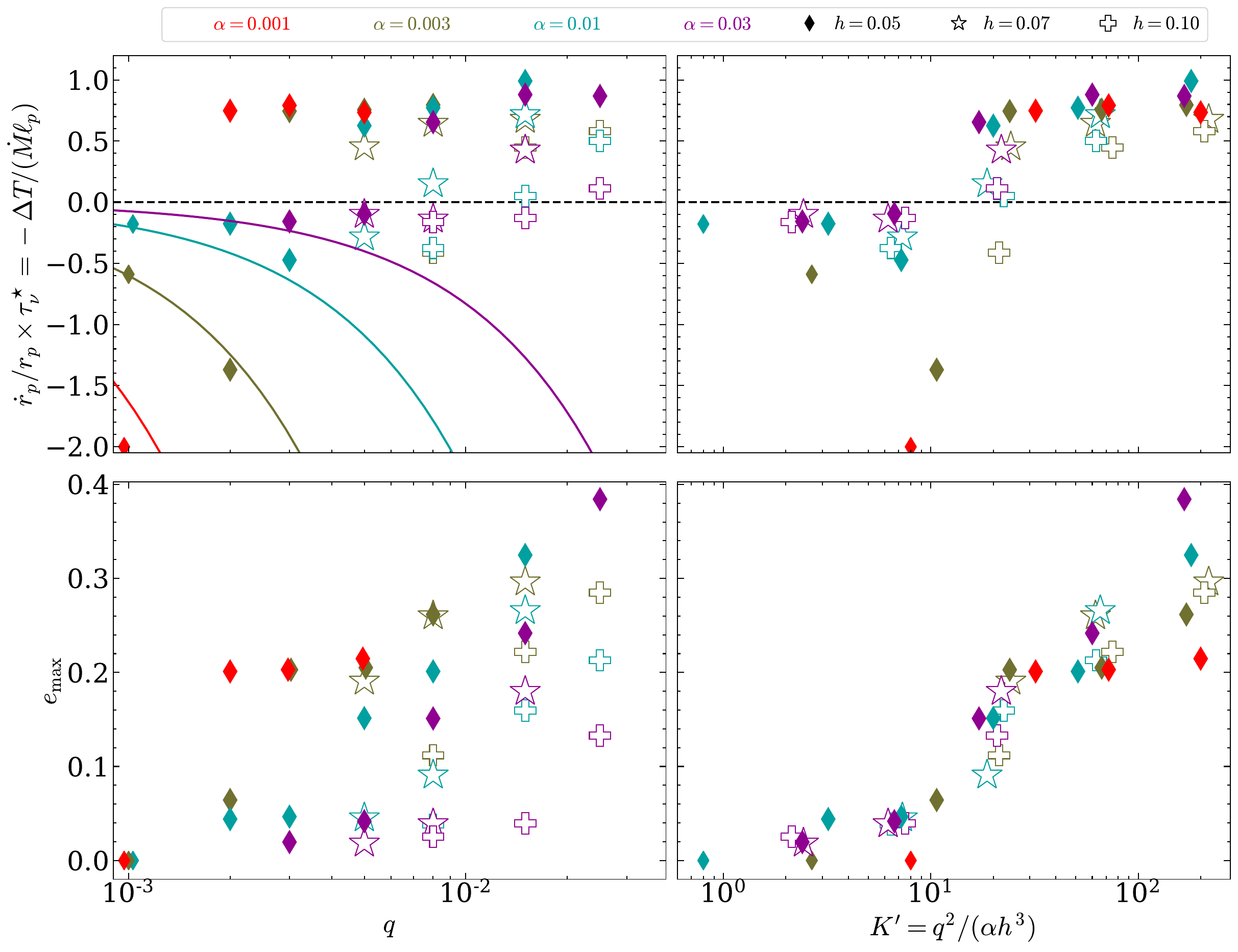} \\
    \caption{
    Dependence of the migration rate and maximum disk eccentricity on $q$ and  gap-width scaling parameter $K'$.  The  curves in the top-left panel summarize  the scalings found from
      \citetalias{DLL} for lower mass ($q\leq 10^{-3}$) planets at $h=0.05$, after extrapolating to these masses. 
      \rev{Note that points with $q=10^{-3}$ are taken from \citetalias{DLL}. }
      }
    \label{fig:summary}
\end{figure*}

The migration rate is obtained by angular momentum conservation,
${1\over 2}M_p \ell_p \dot{r}_p/r_p=-\Delta T$, where $M_p$ is planet mass. 
We may rewrite this as
\begin{equation}
\frac{\dot{r}_{\rm p}}{r_{\rm p}}
 = - {1\over \tau_\nu^*}\left(\frac{\Delta T}{\dot{M} \ell_p} \right) ,
 \label{eq:rdoty}
\end{equation}
where 
\be 
{1\over \tau_{\nu}^*}\equiv {M_d\over M_p}{1\over \tau_\nu}
\ee
is the ``mass-reduced viscous rate,''  which we define in terms of the 
following quantities: the viscous time
$\tau_{\nu}=\tfrac{2}{3} r_p^2/\nu(r_p)$, 
and 
  the proxy disk mass 
$M_d = 4 \pi r_p^2 \Sigma_Z(r_p)$, where $\Sigma_Z = \dot{M}/(3 \pi \nu)$ is the surface density profile in the absence of the planet \citepalias{DLL}.
For comparison,  the   Type II   
migration rate
for planets less massive than the disk
is $-1/\tau_{\nu}$  \citep{Ward.1997}.
And for planets more massive than the disk (upon which we focus), Type II predicts the rate to be $ -(M_d/M_p)^\gamma/\tau_\nu^* $, where $\gamma$
is a positive number whose value depends on the assumed background disk profile
\citep{Syer.1995,Ivanov.1999}.

Figure \ref{fig:summary} (top left panel) shows the 
migration rates from the simulations.
 The curves in the top left panel are from 
 \citetalias{DLL}, in which we ran simulations with $q\leq 10^{-3}$ and $h=0.05$, 
 and fit the resulting $\Delta T$'s  to a power-law expression. 
The extrapolation of those curves to higher $q$ is clearly discrepant 
with our simulations here.

For our present simulations with higher-mass planets, we see that at high enough $q$ planets
migrate outward. For example,
focusing first on the simulation set
with $h=0.05$ (solid diamonds), we see that the subset  with $\alpha=0.01$ transitions
to outward migration at $q\gtrsim 4\times 10^{-3}$. At lower $\alpha$ (and still $h=0.05$), the transition to outward
migration occurs at a lower $q$.

The transition from inward to outward
migration  happens alongside another transition: the disk becomes eccentric. 
In the lower-left panel of Figure \ref{fig:summary}, we 
plot
 the maximum eccentricity in the disk $e_{\rm max}$ vs $q$, where the eccentricity is measured
 by averaging the eccentricity vector within rings, and then time-averaging the resulting (scalar) 
 eccentricity \citep{Teyssandier.2017}.
 Comparing each subset of simulations in the top-left and bottom-left panels, 
 we see that migration is inward in nearly circular disks, and outward in 
 eccentric ones (i.e., when $e_{\rm max} \gtrsim 0.1$).

As seen in Figure \ref{fig:summary},
both transitions occur above a critical $q$ that depends on $\alpha$ and $h$.
We find empirically that the transition occurs at roughly a fixed
value of 
\begin{equation}\label{eq:kprime}
K'\equiv K h^2 = {q^2\over\alpha h^3} \ ,
\end{equation}
The right panels of Figure \ref{fig:summary} re-plot the data in the left panels, but now with $K'$
on the x-axis.  
We observe that at $K'\gtrsim 20$, 
i.e., at
\be
q \gtrsim  1.5\times 10^{-3}\left( {\alpha\over 0.001}  \right)^{1/2}\left( {h\over 0.05} \right)^{3/2}
\label{eq:qlim}
\ee
the migration rate transitions from inward
to outward, and the disk eccentricity transitions from $\lesssim 0.1$ to $\gtrsim 0.2$.  
The parameter $K'$ has been found empirically to correlate with the width of the gap for planets not massive enough to excite a significant disk eccentricity  (\citealt{Kanagawa.2016}; \citetalias{DLL}).
 \citet{Kley.2006} and \citet{Teyssandier.2017} find that for $\alpha=0.004$ and $h=0.05$, disks become eccentric when $q\gtrsim 3 \times 10^{-3}$, in agreement with
Equation \eqref{eq:qlim}.

 The theory laid out in \citet{Lubow.1999} and \citet{Teyssandier.2016,Teyssandier.2017} shows that the disk's eccentricity excitation is sensitive to the density profile, 
 because the density profile controls whether the resonances that excite eccentricity
 are stronger than those that damp it.  Hence the fact that $K'$---and hence gap width---appears to control
 whether disks are eccentric or not is not too surprising.
 Nonetheless, although the theory  of \citet{Teyssandier.2016} successfully predicts many aspects of their simulations, it does not reproduce the correct threshold for where eccentricity is excited, and so  they may
 be missing an effect.

\subsection{Why do planets migrate outward in eccentric disks?}
\label{sec:why}

\begin{figure*}[t]
    \centering
    \includegraphics[trim={0.3cm 0.3cm 0.22cm 0},clip,width=.98\textwidth]{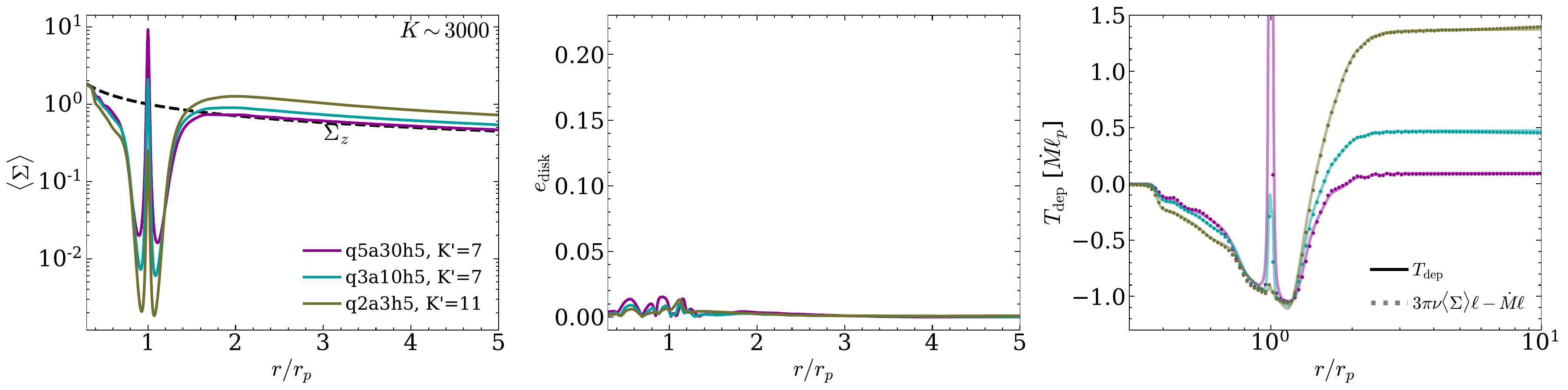}
    \includegraphics[trim={0.3cm 0.3cm 0.22cm 0},clip,width=.98\textwidth]{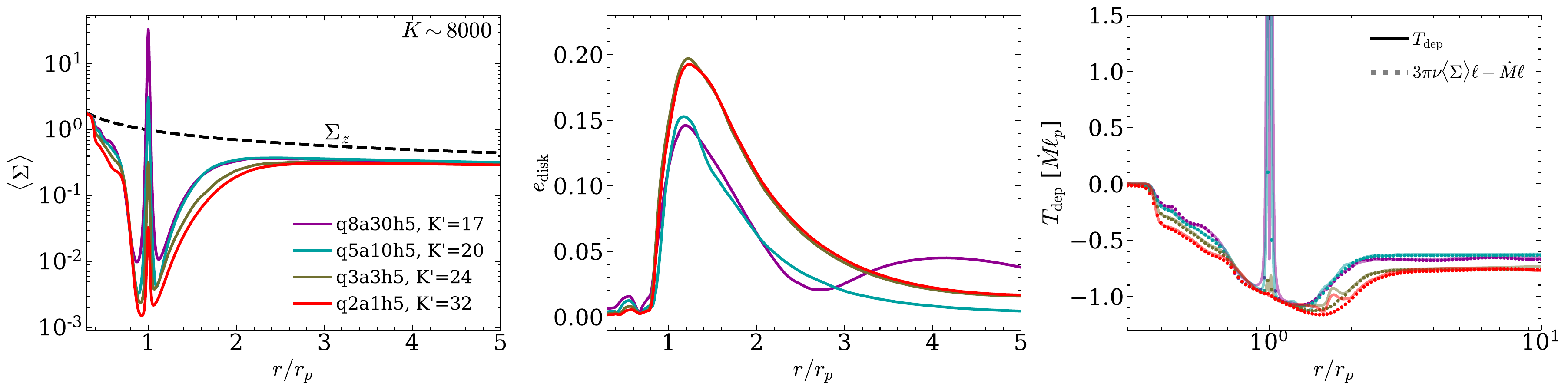}
    \caption{
    Profiles of surface density, eccentricity, and (cumulative) deposited torque in some nearly circular disks (top panels) and eccentric disks (bottom).
    \rev{Each profile is time-averaged over the final $10,000$ orbits of the simulation, after the torque has reached its steady-state value.}
   Eccentric disks have more extended outer gaps and weaker outer torques.  The latter forces their total torque, $\Delta T\equiv T_{\rm dep}\vert_{r\gg r_p}$, to be negative.
   Simulations are labeled such that e.g., "q8a30h5" corresponds to $q=8\times10^{-3}$, $\alpha=30\times10^{-3}$, and $h=5 \times 10^{-2}$.
   }
    \label{fig:profs}
\end{figure*}

Figure~\ref{fig:profs} shows the steady-state profiles of surface density and eccentricity
for simulations that have $h=0.05$ and $K'$ near the migration transition.
In the top row, migration is inward and the disk is nearly circular, and in the bottom, migration is outward and the disk is eccentric.
The profiles are
time-averaged in steady-state over 10,000 orbits, which is much longer than the precessional times.
As seen in the $\Sigma$ profiles, higher values of $K'$ systematically  result in deeper gaps.

In the circular disks, the gaps are nearly symmetric relative to the planet's position, while
in the eccentric disks, the outer half of the gap becomes significantly wider than the inner one.
In addition, the circular disks have a modest density pile-up outside of the planet's orbit, and the eccentric
ones have a  deficit.  The sign and magnitude of the pileup is dictated
by the value of $\Delta T$ 
(as shown in Equation 19 of \citetalias{DLL}, which follows from Equation~\ref{eq:angflux} below).

Turning to the eccentricity profiles, we see that 
above the transitional $K'$  it is the outer disk that develops a significant
eccentricity. 
And although not shown in the figure, the eccentric
disks precess coherently,\footnote{For some large $q$ simulations, we find a second bump in the  eccentricity profile  in the outer disk that has a low amplitude (e.g., the purple curve in the lower middle panel of Figure \ref{fig:profs}), and that precesses at a different
rate than the main bump.
But as seen in the right panel, the torque is unaffected by this secondary mode.
} and the amplitude of the eccentricity
remains unchanged over multiple viscous times. 
This points to the eccentricity being a free mode of the disk \citep{Teyssandier.2016,Lee.2019}, with an excitation that balances viscous damping to maintain the mode in steady-state \citep[as in circumbinary disks; see][]{Munoz.2020b}.

The right panels 
 show  $T_{\rm dep}$, the
 cumulative
 torque deposited into the disk
 (within each $r$) by the damping of waves.
 To calculate $T_{\rm dep}$, we evaluate the
angular momentum flux carried by waves, and then remove from that the gravitational torque excited  by the planet \citepalias{DLL}.
We see that throughout most of the outer disk   $T_{\rm dep}$ increases with $r$, showing that the torque deposition per unit $r$ is positive there; i.e., that region is responsible for pushing the planet inward. Similarly, most of the inner disk pushes the planet outward, as $T_{\rm dep}$ decreases with $r$ when $r<r_p$. Far beyond the planet---
 outside a ${\rm few} \times r_p$---$T_{\rm dep}$ flattens and approaches $\Delta T$,
i.e, all of the torque excited by the planet is eventually deposited in the disk.
We  henceforth  use the $T_{\rm dep}$ profiles to examine what is causing $\Delta T$ to become negative
in  eccentric disks.

Figure \ref{fig:profs} shows that the torque on the inner disk is always
\be 
T_{\rm dep}(r=r_p)\approx-\dot{M}\ell_p \ .\label{eq:minusone}
\ee
Equation \eqref{eq:minusone} is a generic result for deep gaps in steady-state disks. It follows 
from  the conservation of angular momentum flux, which we approximate here as
\begin{equation}
T_{\rm dep}(r)\approx 3\pi\nu\langle\Sigma\rangle \ell-\dot{M}\ell   , \label{eq:angflux}
\end{equation}
(see \citetalias{DLL} for the full expression).
Therefore if the gap is sufficiently deep, 
the first term on the right-hand side may be neglected at $r_p$, 
confirming Equation \eqref{eq:minusone}.
In the right panels, we also convert the $\Sigma$ profiles to $T_{\rm dep}$
via Equation~(\ref{eq:angflux}).  The result is shown as \rev{dotted} curves. The fact that these agree with 
the directly-calculated $T_{\rm dep}$ confirms that the disk is in viscous steady state.

\begin{figure}
    \centering
    \includegraphics[trim={0.3cm 0.3cm 0.22cm 0},clip,width=.48\textwidth]{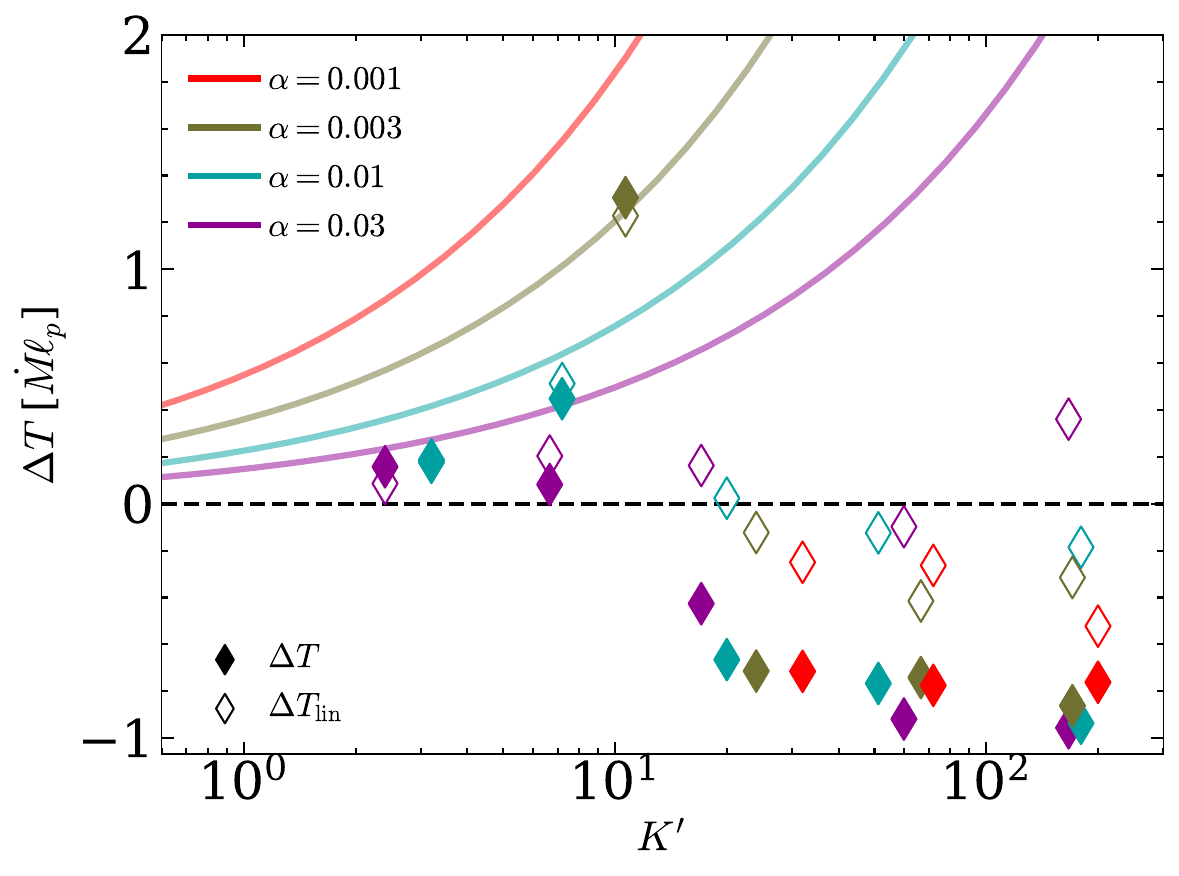}
    \caption{ 
    Comparison of $\Delta T$ (filled symbols)
    to the linear $\Delta T$ (open symbols) calculated from the $\avg{\Sigma}$ profiles, for disks with $h=0.05$.
    The solid lines show the extrapolations of the circular disk results of \citetalias{DLL} to these masses. 
    }
    \label{fig:linear}
\end{figure}

Given the value of the torque on the inner disk (Equation \ref{eq:minusone}), it  must be that 
$\Delta T$ becomes negative in  eccentric disks because the torque
on the outer disk weakens.
But why?
From the theory in Section \ref{sec:pd}, we must examine torque excitation and deposition. 
\begin{enumerate}[(a)]
\item  Excitation: 
For a given $\Sigma$ profile, 
eccentricity could lower 
$\Delta T$ 
if it
lowers
the amplitude of the excited waves in the outer disk, and hence lowers the angular momentum carried by them.  (We also include in this category the change in co-orbital torques; see below.)
\item   Deposition: 
for a given $\Delta T$, 
eccentricity  could increase the distance  waves travel
 before they damp. 
A longer damping length means
a wider gap (from Equation~\ref{eq:angflux}), which implies that at the location where the waves
are excited $\Sigma$ is lower, and hence the wave amplitude is lower.
\end{enumerate}

We examine  (a) in Figure \ref{fig:linear} by
 comparing  the simulated values of $\Delta T$ 
 with a linear calculation of $\Delta T$
 in a  circular disk whose $\Sigma(r)$ profile is the same as from the simulation\footnote{\rev{One could check hypothesis (a) directly by extending the linear theory of torque excitation to an eccentric disk.
 We defer such an analysis to future work.}
 } (see \citetalias{DLL} for our linear calculation method.)
We see that for the circular disks ($K'\lesssim 20$), the linear calculation adequately predicts what is found in the simulations. 
But for the eccentric disks it overpredicts $\Delta T$.
We infer that the disk being eccentric lowers $\Delta T$, 
as suspected in (a). 
A more detailed examination  shows
that much of the lowering is indeed due to outer  waves carrying less angular momentum
in the simulated eccentric disk (relative to the linear calculation). 
But there is an additional effect: 
the co-orbital torques
 become stronger (more negative) in the eccentric disks. 
 We find this effect to be subdominant up to $K'\sim 50$; for $K'$ larger than that, 
 the two effects are comparable.

 Turning to (b),  the right panels of Figure \ref{fig:profs} show that the damping length is indeed longer in the
eccentric disks.  Unfortunately, we have not been able
to quantify the relative importance of (a) and (b). 
For a partial quantification of the latter, the
 curves in Figure \ref{fig:linear} show the  extrapolations from  the circular-planet simulations (as in Figure \ref{fig:summary}). 
 These lie above the open diamonds 
 for the eccentric disks
($K'\gtrsim 10$), and one of the reasons for the discrepancy 
is that the gap shape for the open diamonds is wider than would be implied
in the extrapolations. 
In truth,  explanations (a) and (b) are entangled:
e.g., when waves are excited to lower amplitudes they travel further before becoming nonlinear and damping. But for
a first-principles derivation, one must extend the theory of 
\cite{Goodman.2001} to investigate  more massive planets and viscous disks.

\section{Summary and Discussion}
\label{sec:sum}

\subsection{Summary}
Figure \ref{fig:summary} encapsulates our main result:
planet migration transitions from inward to outward when the
planet exceeds
around two Jupiter masses, for the fiducial values 
$\alpha=0.001$ and $h=0.05$  (Equation~\ref{eq:qlim}).
This transition coincides with 
the  disk becoming eccentric. 
In Section \ref{sec:why} we showed  why eccentric disks lead
to outward migration.

We find that the total torques are  always close in magnitude to the advection  torque
($|\Delta T|\sim \dot{M}\ell_p$), whether migration is inward or outward
(Figure \ref{fig:summary}). 
As a result, the 
migration rates are  comparable in magnitude to the ``mass-reduced viscous rate''
(Equation~\ref{eq:rdoty}).\footnote{ 
For comparison, 
 the Type II rate (given below Equation~ \ref{eq:rdoty}) is dramatically slower than the mass-reduced viscous rate
 when $M_d\ll M_p$. But for $M_d\sim M_p$, the  Type II rate is close to the true rate in magnitude---even though Type II  can predict the wrong sign. 
}
It is  not  surprising 
that $|\Delta T|\sim \dot{M}\ell_p$
when migration is outward, because the inner 
torque is always equal to  $-\dot{M}\ell_p$ for deep gaps (Section \ref{sec:why}); and so if the outer torque is subdominant, the total  torque will also be close to $-\dot{M}\ell_p$. But 
it is surprising  that, even when migration is inward,
it is never the case that $\Delta T\gg \dot{M}\ell_p$. 
We speculate 
that
this is because a planet always acts as a leaky sieve, and so 
the pileup beyond the planet can never be too big.  
That would in turn imply that the normalized torque cannot be too large, because of the close connection between the pileup and the normalized
torque
 \citepalias{DLL}.
But why the planet should always
act as a leaky sieve remains an open question.

\subsection{Assessment of Key Assumptions}

 \begin{itemize}

    \item $\alpha$-model: Perhaps our most questionable assumption
    is that of $\alpha$-viscosity.
    If angular momentum 
    is transported by a non-viscous mechanism, such as disk winds, then the gap
    shapes would be different, and hence so would the torques.

     \item  low-mass disk: Fixing the planet's orbit is a good
    approximation when the planet's migration timescale ($\sim \tau_\nu^*$)
    is shorter
    than the gas radial drift
    timescale  ($\tau_{\nu}$), or equivalently, when $M_d\lesssim M_p$.

\item circular planet orbit: 
An eccentric planet will introduce a forced component
to the disk eccentricity, on top of the free component that is excited above the transitional mass.
The planet's eccentricity will also
modify the torque on the disk.
\rev{\cite{DAngelo.2006}  include the backreaction of the disk 
on the planet, and find that the planet's eccentricity is rapidly excited, 
which affects the migration direction.
But their inner disks are severely depleted, perhaps due to their inner boundary condition, an insufficient run time compared to the viscous time, or to material being lost onto the planet.
Moreover, their disks are at least as massive as the planet, invalidating our low-mass disk assumption.
In the future, we plan to determine whether the planet's eccentricity is excited or damped in low-mass steady disks, as has been done for stellar binaries \citep[e.g.,][]{Munoz.2019,Dorazio.2021}.
In the latter case, it has been found that the binary's eccentricity damps if it is below a threshold eccentricity of around 0.1, even though the disk is very eccentric.
}

   \item  2D disk: For high-mass planets, gaps are much wider
    than the disk scale height, and so the 2D treatment should  
    adequately capture the excitation and deposition torques.

    \item  no accretion onto planet: 
    Our large softening parameter prevented planetary accretion.
    Realistic modeling of accretion requires high resolution, in addition to
    a more accurate treatment of the thermal state of the gas, \rev{disk self-gravity},
    and three dimensions
         \citep[e.g.,][]{Fung2019}.
     If a significant fraction
     of the disk's radial mass flow
     ends up in the planet, the migration rates would change significantly.
     \rev{
     Additionally, the torque from the circumplanetary region is unrealistic.
     But we have checked that this torque is small,
     typically contributing between $0.1-0.25$ to the dimensionless migration rates in Figure \ref{fig:summary}.}

 \item 
 locally isothermal equation of state: 
 Adopting a finite cooling time, rather than the instantaneous cooling we that we assume,
  would affect the deposition profile.
  That is because
 linear waves  deposit angular momentum into the disk even
 in the absence of viscous dissipation, and the
 amount depends on the cooling time \citep{2020ApJ...892...65M}.
 We suspect this dependence is minimal for the very massive planets considered in this Letter, for which viscous dissipation (including by shocks)
 likely dominates over the aforementioned linear wave deposition.
 
 \item \rev{inner boundary condition:  For our results not to depend on the inner boundary,
the inner wave-killing zone must be far enough from the planet that most of the torque excitation occurs
   within the simulation domain \citepalias{DLL}. 
 We have checked that by examining the excitation profile. 
In addition, we have tested roughly half of our $h<0.1$ simulations by continuing them in a domain with a smaller wave-killing boundary (down to $0.2 r_p$), and found in these cases that the change in torque was small. Note that the $h=0.1$ simulations require a larger radial domain (see footnote \ref{foot:2}), because the  wave excitation profile is broader at larger $h$.}

\end{itemize}

\subsection{Comparison to Prior Work}
Previous studies have found outward migration of massive planets, but these require either steeply falling surface density profiles \citep{Chen.2020}, or extremely massive disks to induce either Type III migration \citep{Masset.2003} or gravitational instability \citep{Lin.2012,Cloutier.2013}. 
 By contrast, we find outward migration more generally for super-Jupiters, subject to the assumptions listed above.

Recently, \citet{Duffell.2020} examined a setup  similar
to ours, and found that $\Delta T$ reverses sign in the brown dwarf regime, but for $h=0.1$ and $\alpha\ge0.03$. 
Using Equation \eqref{eq:qlim} for their parameters, we find general agreement with their transitional masses (cf. their Figure 5), suggesting that the torque reversal reported by \citet{Duffell.2020} is equivalent to ours.

\subsection{Directly Imaged Planets}

The two-Jupiter-mass threshold that we have uncovered is intriguingly similar to the lowest mass giant planets discovered via direct imaging techniques \citep[e.g.,][]{Bowler.2018}. 
 Due to their large masses and wide separations, these planets were once thought to originate from gravitational instabilities. But recent evidence points to
 a mass distribution that may be consistent with the core accretion 
 scenario \citep{Nielsen.2019,Wagner.2019}. Thus,
  those super-Jupiters detected via direct imaging could be the outcome
 of core accretion, but followed by outward migration in an eccentric disk.
This scenario would not require the presence of additional giants, as would be the case for jointly migrating planets trapped in mean motion resonance \citep{Crida.2009}.
It has indeed been speculated that outward migration could explain the properties of directly imaged planets, e.g., as pointed out recently by \citet{Bohn.2021} in the context of their discovery of a $\sim 6$ Jupiter mass planet at $\gtrsim 115$ AU from its star. 
We propose that such planets could have migrated via interaction with an eccentric disk.

\acknowledgements

We than W. Kley for helpful comments, and the anonymous referee for a careful report that improved the quality of the paper.
A.M.D gratefully acknowledge the support by LANL / LDRD and NASA /  ATP. 
This research used resources provided by the Los Alamos National Laboratory Institutional Computing Program, which is supported by the U.S. Department of Energy National Nuclear Security Administration under Contract No. 89233218CNA000001.
The LANL report number is LA-UR-20-29630. 
This work used computing resources at CIERA funded by NSF PHY-1726951.
D.J.M. acknowledges support from the CIERA Fellowship at Northwestern University and the Cottrell Fellowship Award from the Research Corporation for Science Advancement, which is partially funded by the NSF grant CHE-2039044.
{Y.L. acknowledges NASA grant NNX14AD21G and NSF grant AST-1352369.

\begin{appendix}
\section{The indirect potential} \label{sec:app_ind}

We run our simulations in the stellocentric frame, where the gravitational acceleration for a fluid element located at position $\vec{r}$ relative to the star is, 
\be \label{eq:app_ad}
\ddot{\vec{r}} &=& -G M_\star \frac{\vec{r}}{r^3} -  G M_p \frac{(\vec{r}-\vec{r}_p)}{|\vec{r}-\vec{r}_p|^3}  - G M_p \frac{\vec{r}_p}{r_p^3}  - G \int \frac{\vec{r}}{r^3} dm ,
\ee
 where $dm = \Sigma r dr d\phi$ is the mass element of the disk.
The first two terms are the direct accelerations due to the star and planet, while the last two terms are the indirect accelerations due to the acceleration of the star by the planet and disk.
We neglect the last term in the simulations both
because we consider small disk masses, and because it
 is proportional to disk eccentricity, and so averages out on the precessional timescale.

We now consider how the indirect terms
affect the torque measurements.
We will show that the last term in Equation \eqref{eq:app_ad} has a subdominant effect, and so we do not include it in our torque measurements in the body of the paper.
Equation \eqref{eq:app_ad} is evaluated in the stellocentric frame, but the total torque on the planet-star binary is most easily evaluated in the barycentric frame. 
In this frame, the disk torques the binary according to 
\be \label{eq:app_lb}
\dot{L}_b = \int \left[ GM_\star \frac{  \vec{R} \times \vec{r}}{r^3} +G M_p \frac{ \vec{R} \times (\vec{r}-\vec{r}_p)}{|\vec{r}-\vec{r}_p|^3} \right] dm ,
\ee
where $L_b$ is the angular momentum of the binary and $\vec{R}$ is the position of a disk element with respect to the star-planet-disk center of mass.  
Note that the third term of Equation \eqref{eq:app_ad}, which is the acceleration of the star due to the planet, does not contribute to $\dot{L}_b$.
The stellocentric and barycentric disk positions are related via the center of mass position of the star, $\vec{r}= \vec{R}-\vec{R}_\star$, where,
\be
\vec{R}_\star = - \mu_p \vec{r}_p - \mu_d \vec{r}_d ,
\ee
and where $\mu_{p,d} = M_{p,d}/(M_\star + M_p + M_d)$ are the mass ratios of the planet and disk, $M_d = \int dm$ is the disk mass, and $\vec{r}_d = \int \vec{r} dm/M_d$ is the center of mass position of the disk.
Carrying out the cross products in terms of the stellocentric positions and approximating $\dot{L}_b \approx \dot{L}_p = M_p \ell_p \dot{r}_p/(2 r_p)$
for small mass ratios we find,
\be \label{eq:app_lbdot}
\dot{L}_p &\approx& G M_p \int \vec{r} \times \frac{\vec{r}-\vec{r}_p}{|\vec{r}-\vec{r}_p|^3} dm  \\
            &-& G M_p \int \vec{r}_p \times  \frac{\vec{r}}{r^3}  dm \nonumber \\
            &-& G M_d \int \vec{r}_d \times \left[ \frac{\vec{r}}{r^3} + q \frac{\vec{r}-\vec{r}_p}{|\vec{r}-\vec{r}_p|^3} \right] dm . \nonumber            
\ee
The first line is the familiar torque of the disk on the planet due to the direct acceleration (second term of Equation \ref{eq:app_ad}). 
The second and third lines are corrections arising from the displaced disk center of mass (the last term in Equation \ref{eq:app_ad}). 
We neglect the contribution from the third line as we find that $\vec{r}_d$ in all simulations is on the order of $e_{\rm max} r_p$, making the third line an order $e_{\rm max} M_d/M_p \ll 1$ correction.
The remaining correction to the direct torque is the second line, which we measure, {\it a posteriori}, to be $\lesssim 0.2 \dot{M} \ell_p$. 
If we were to include this term in the $\Delta T$ values shown in Figure \ref{fig:summary} (which only contain the direct torque), the points at $K\gtrsim20$ would shift by a minor amount, but not enough to affect our main conclusions.

\section{Barycentric outer boundary condition} \label{sec:app_bc}

We apply the inflow boundary condition of \citetalias{DLL} at the disk's outer boundary. 
The distant inward flow  is
assumed to be axisymmetric around the star-planet barycenter, which is displaced from the origin of the simulation.
We prescribe the inflow rate, $\dot{M}$, and set $\Sigma = \dot{M}/(3\pi\nu)$ at the boundary.
The fluid velocities are set to their barycentric values: $v_{r,{\rm bary}} = \dot{M}/(-2\pi r_b \Sigma)$, where $r_b$ is the distance to the barycenter, and $v_{\phi,{\rm bary}}$ is the pressure corrected Keplerian velocity.
We then transform the barycentric velocities to \fargo's stellocentric, rotating frame using
\be \label{eq:app_rot}
\begin{pmatrix}
v_r  \\
v_\phi
\end{pmatrix}
=
\begin{pmatrix}
\cos(\phi_b - \phi) & -\sin(\phi_b -\phi) \\ 
\sin(\phi_b - \phi) & \cos(\phi_b -\phi) \\ 
\end{pmatrix}
\begin{pmatrix}
v_{r,{\rm bary}}  \\
v_{\phi,{\rm bary}} - n r_b
\end{pmatrix} ,
\ee
where $n=\sqrt{1+q}\Omega_p$ is the planet's mean motion, and $\phi_b$ is the azimuthal angle in the barycentric coordinate system.
For each cell along the stellocentric boundary with Cartesian stellocentric coordinates $(x,y)$, we first compute $r_b = \sqrt{(x-\mu r_p)^2 + y^2}$, where $\mu = q/(1+q)$, and then apply the transformation given by Equation \eqref{eq:app_rot} with
\be
\cos(\phi_b - \phi) = \frac{r}{r_b} \left(1 - \frac{\mu r_p}{r} \right) , \qquad \sin (\phi_b - \phi) = \frac{\mu r_p}{r_b} .
\ee
\citet{Teyssandier.2017} used a boundary condition similar to Equation \eqref{eq:app_rot}, but with $v_{r,{\rm bary}}=0$.

\section{Convergence of simulations} \label{sec:app_conv}

\begin{figure*}[t]
    \centering
    \includegraphics[trim={0.25cm 0.25cm 0.25cm 0},clip,width=.98\textwidth]{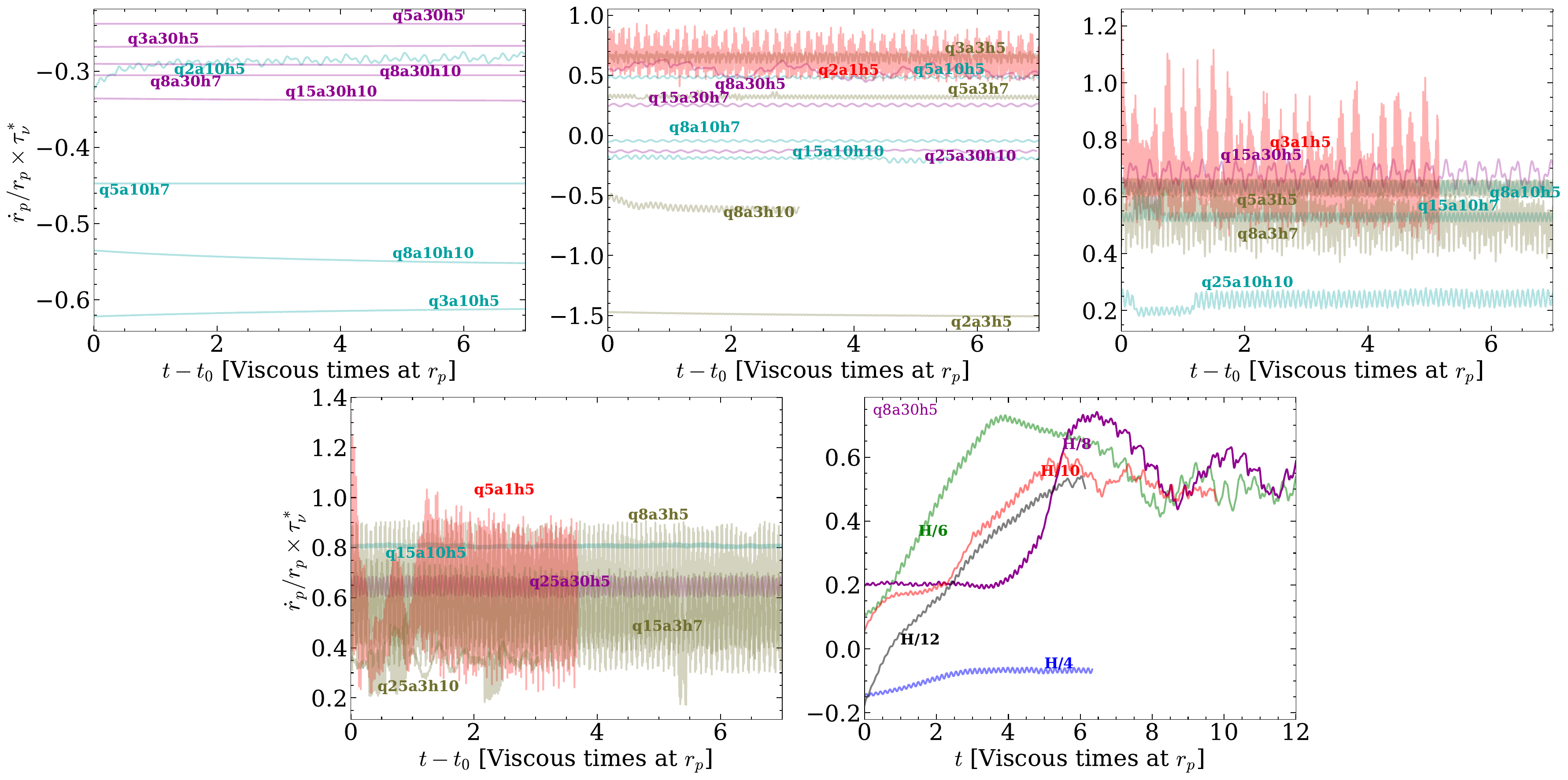}
    \caption{
    Convergence of  migration rates with time (first four panels) and with resolution for one example simulation (last panel).
    \rev{In the first four panels,} we show only the final 7 viscous times, \rev{while in the last we show the first 12 viscous times}.
    For each simulation we time average each point over $1000$ orbits to remove short timescale variations.
    See the caption of Figure \ref{fig:profs} for our naming convention.
    }
    \label{fig:conv}
\end{figure*}

We run each simulation in two stages. 
In the first stage, \rev{we start the simulation with the steady-state}
\rev{planet-less solution,} 
fix $\dot{M}$ at both boundaries, and evolve the disk long enough for the gap to open and reach a quasi-steady-state. 
During the second stage, we switch to the inner boundary condition described in Section \ref{sec:bc} that allows $\dot{M}$ at the inner boundary to adjust as the disk evolves to a global steady-state.

Simulations reach steady-state once the time- and azimuthally-averaged $\dot{M}$ is independent of $r$. 
For our lowest $\alpha$'s, we find that the average $\dot{M}$ in the outer disk ($r>1.05 r_p$) is within $10\%$ of the average inner disk $\dot{M}$ ($r<1.05 r_p$).
These $\dot{M}$'s typically agree to less than $1\%$ in the larger $\alpha$ cases.

In the first four panels of Figure \ref{fig:conv} we show that all of our migration rates are converged in time. 
Each point is normalized to the current value of $\dot{M}$ at the inner boundary and time averaged over a window of 1,000 orbits to remove short timescale oscillations.
For clarity, we only show the final $7$ viscous times, but note that many of the simulations are converged in the first few viscous times.

The final panel of Figure \ref{fig:conv} shows the dependence of $\Delta T$ on the simulation resolution for one exemplary simulation with $q=0.008$, $\alpha=0.03$, and $h=0.05$.
We find that once the resolution exceeds $6$ points per scale height, the migration rate is converged. 
Moreover, we showed in \citetalias{DLL} with many more simulations that our fiducial resolution of 8 points per scale height was adequate for determining $\Delta T$.

As discussed in Section \ref{sec:bc}, some simulations use a smaller inner boundary in order to capture all of the torque excited by the planet. 
The amount of missing torque is typically small for simulations with $h<0.1$ that have an inner boundary at $0.3 r_p$.  
The only exception to this is the q8a30h5 simulation, for which moving the inner boundary to $0.1 r_p$ lowered the migration rate by $\sim 50\%$. 
This reduction was due to a $\sim 50\%$ smaller disk eccentricity, which resulted in a comparatively larger outer disk torque.
Nonetheless, this simulation was on the boundary between low and high eccentricity (see Figure \ref{fig:summary}), and therefore does not affect the transitional $K'$ that we find.

\end{appendix}

\end{document}